\documentclass[11 pt]{article}
\usepackage[margin=1.2in]{geometry}

\usepackage[utf8]{inputenc}
\usepackage{amsmath} 
\usepackage{amsthm} 
\usepackage{amssymb}
\usepackage{mathrsfs}
\usepackage{bbm}
\usepackage{bm}
\usepackage{bookmark}
\usepackage{hyperref}
\usepackage{algorithm}
\usepackage{algorithmicx}
\usepackage{algpseudocode}
\usepackage{graphicx}
\usepackage{subcaption}
\usepackage{xcolor}

\newtheorem{theorem}{Theorem}
\newtheorem{corollary}{Corollary}

\newtheorem{proposition}{Prop.}

\newtheorem{remark}{Remark}
\newtheorem{assumption}{Assumption}

\DeclareMathOperator{\EX}{\mathbb{E}}
\DeclareMathOperator{\Ind}{\mathbbm{1}}

\title{Optimal Intervention in Economic Networks using Influence Maximization Methods}
\author{Ariah Klages-Mundt\\Cornell University, Center for Applied Mathematics \and 
	 Andreea Minca\\ Cornell University,  School of Operations Research and Information Engineering
	 }

\date{October 11, 2021}

\begin{document}

\maketitle

%
%
%
%
%
%
%
%
%
%
\begin{abstract}
We consider optimal intervention in the Elliott-Golub-Jackson network model \cite{jackson14} and we show that it can be transformed into an influence maximization-like form, interpreted as the reverse of a default cascade. Our analysis of the optimal intervention problem extends well-established targeting results to the economic network setting, which requires additional theoretical steps. We prove several results about optimal intervention: it is NP-hard and cannot be approximated to a constant factor in polynomial time. In turn, we show that randomizing failure thresholds leads to a version of the problem which is monotone submodular, for which existing powerful approximations in polynomial time can be applied. In addition to optimal intervention, we also show practical consequences of our analysis to other economic network problems: (1) it is computationally hard to calculate expected values in the economic network, and (2) influence maximization algorithms can enable efficient importance sampling and stress testing of large failure scenarios. We illustrate our results on a network of firms connected through input-output linkages inferred from the World Input Output Database.

\end{abstract}

\section{Introduction}
Following the global crisis due to the COVID-19 medical and economic contagion,  governments have unleashed unprecedented macroeconomic stimulus. The variety of proposed stimulus, both in government financing and in monetary policy form, aims to support value in a shocked global economy.
The tools to support value following a systemic shock are there since the financial crisis, and new ones are being proposed.  One  difference to the financial crisis is that the shock originated then from within the financial system and the main intervention target were systemically important  institutions, i.e., those whose failure would lead to a large impact on the  economy. 
In this crisis the shock was external and created disruptions to many economic sectors worldwide. Consequently, intervention is much more widespread.

As learned from the financial crisis, network effects underpin systemic importance, which can be measured   based on the size of loss cascades, see e.g, \cite{amini2016resilience, detering2019managing} or centrality measures, see \cite{bartesaghi2019risk} and the references therein.  Work on systemic risk measures, e.g.,\cite{chen13,biagini19,feinstein2017measures,Armenti}, led to different axiomatic frameworks for capital requirements such that aggregate risk is acceptable. Notably,  aggregation functions underlying these systemic risk measures can account for interconnections.
In \cite{amini2015control,amini2017optimal,capponi2015systemic}, authors explore  optimal capital and liquidity intervention, and derive insights into the intervention target in stylized core-periphery banking networks subject to the risk of bank runs. Their methods are applied for small banking systems.  In \cite{AHN2019257}, authors cast the intervention  problem in the context of the Eisenberg-Noe model \cite{eis} as a mixed integer-programming problem, and propose a notion of of $\epsilon$-optimality to solve it approximately. They apply their methods to the Korean banking system.
In contrast to these past works, our paper focuses on the computational aspect of  optimal intervention problems, which becomes critical when the number of eligible firms is large. 
When entire sectors, rather than a few large institutions, are hit by shocks, one  needs to understand the systemic impact of groups of firms  and optimally decide on where to intervene. Such problem quickly becomes computationally hard.
The government's criterion is to maximize the overall value in the system under a budget constraint.

Our model relies on the notion of value of an organization --firm, sector, country-- introduced in \cite{jackson14} in the context of cross-holdings. Without intervention, if the value of the organization drops below a failure threshold, then there are 
failure losses and  the values of the connected organizations drop as well and so on. This is also in the spirit of the distress notion in \cite{veraart20}, which allows for contagion before the point of default. The failure threshold is interpreted as the value below which the organization ceases operations.  Intervention can be seen as a way to increase an organization's value or alternatively lower its failure threshold.
Several types of interventions can be modeled by a decrease of the failure threshold of an organization.
Government bailouts could take the form of  equity infusions, as they did in the financial crisis.  Central banks are injecting liquidity in the economy via various asset purchase programs, including corporate debt purchases.

It is clear that direct government financing allows firms to survive by directly lowering the failure threshold. The effect of  asset purchase programs (APP) is more subtle.
A point of contention is whether asset purchase programs involve  liquidity injection, or whether they involve value injection. When central banks can purchase corporate debt  they change the  outcome in debt markets.\footnote{Arguably, central banks can lower the failure thresholds  even without actual liquidity injection: for example Boeing raised debt in capital markets following the FED's announcement that they would support corporate debt markets, see e.g. \url{thttps://www.bloomberg.com/news/articles/2020-05-02/the-non-bailout-how-the-fed-saved-boeing-without-paying-a-dime}.}
An unavoidable fact of APP is that, whenever the central bank purchases illiquid assets to intervene in liquidity, it must price those assets in some way. Models are usually used to calculate a `fundamental value'. When acting as a lender of  last resort, central banks may essentially accomplish bailout functions.
Our model captures both direct and indirect ways of lowering the failure thresholds,  as the value of the organization increases by the intervention amount.

Interventions may be accompanied by long-term moral hazard effects. Firm default is an important long-term filter that incentivizes strong and competent management. The prospect of intervention can disincentivize proper risk management, enabling additional short-term profits to management and equity holders while transferring tail risks to government. Note, however, that interventions can be shaped to reduce moral hazard (e.g. by organizing bail-ins by the creditors and thereby diluting equity holders).  
In \cite{bernard2017bail}, authors endogenize intervention for a network of banks. In their paper, a bail-in can be organized in equilibrium if and only if the
regulator's no-intervention threat is credible,  namely in the last stage of the game she could optimally abandon intervention.
Our work is complementary and could be used for the last stage of such a game, as we find the organizations that need intervention.
We leave moral hazard considerations for future work, given that the widespread consensus of decision makers was to first preserve value following the COVID-19 crisis.
We focus on the specific question of how to design targeted interventions that exploit network effects while leaving the precise micro structure of those interventions as a separate problem.

Our work is also part of the broader literature on targeting in networks, see e.g., \cite{Ballester,galeotti2009influencing}, and in particular the literature on optimal diffusions of products or innovations or influence maximization, \cite{domingos2001mining,kempe03, kempe2005influential}. Our contributions are summarized below.

\paragraph{This paper.}
We construct an economic network intervention model and show how it can be solved by adapting influence maximization methods (Section~\ref{sec:model}). Our analysis extends well-established targeting results to the economic network setting,  requiring additional theoretical steps over the classical setting. For instance, the dependency matrix (``influence matrix'') is more complex ($\sim$ the Neumann series of the matrix in a linear influence setting that is column-substochastic with zero diagonals) taking into account the effect of a firm on itself and the structure of default reversals (``activations'') is more nuanced. We contribute the following results, which provide the groundwork for adapting powerful targeting algorithms to solve several economic network problems:
\begin{enumerate}
    \item We define an optimal economic network intervention problem and show how it can be expressed in an influence maximization-like form (Section~\ref{sec:opt_int}).
    
	\item We prove that it is NP-hard to optimize the economic network intervention and cannot be approximated to a constant factor in polynomial time (Theorem~\ref{result:interventions_hard} and Corollary~\ref{result:approx_hard}).
	
	\item We prove that, when modified to consider expected values under random thresholds, the intervention problem is monotone submodular (Theorem~\ref{result:intervention_submodular}) and thus admits a greedy polynomial time $(1-1/e-\epsilon)$-approximation (Corollary~\ref{result:greedy_approx}).
	
	\item We show that similar results extend to a related problem: identifying large failure cascade scenarios. We prove that it is NP-hard to find the worst case failure scenarios given a maximum sized aggregate shock to asset values (Theorem~\ref{result:max_shock_hard}). Under randomized thresholds, a similar greedy approximation is applicable.
	
	\item We show two practical consequences of Theorem~\ref{result:max_shock_hard} in Section~\ref{subsec:large_cascades}. (1) It is computationally hard to calculate expected values in the economic network. (2) Intervention approximation algorithms can be applied for importance sampling to identify instances that lead to tail events, which can be very valuable applied to stress testing. The depth of sampling in the tail can be tailored by choosing a parameter.
	
	\item We demonstrate a proof-of-concept of optimal intervention approximation applied to economic networks constructed from the World Input-Output Database (Section~\ref{sec:simulation}).
\end{enumerate}

\section{Model}\label{sec:model}
In this section, we supplement the Elliot-Golub-Jackson  network contagion model  \cite{jackson14} to incorporate targeted interventions. We then formulate an optimal intervention problem that relates the economic intervention problem to influence maximization problems.

\subsection{Financial network contagion model}

We define an economic network $(C,D,\beta, \bm\theta, \mathbf p)$ based on the  Elliot-Golub-Jackson  network contagion model as follows:
\begin{itemize}
	\item $U = \{1,2,\ldots,n\}$ the set of firms/nodes in the network
	\item $m$ assets owned by firms
	\item $\mathbf p = m\times 1$ vector of asset prices
	\item $D = n\times m$ matrix with $D_{ik} \geq 0$ the share of asset $k$ held by firm $i$ (adding to 1)
	\item $C = n\times n$ matrix with $C_{ij} \geq 0$ the fraction of firm $j$ owned by firm $i$ and 0 along the diagonals
	\item $\hat C = n\times n$ diagonal matrix with $\hat C_{ii} = 1 - \sum_j C_{ji}$ the share of organization $i$ not owned by another firm in the system
	\item $\bm\theta = n\times 1$ vector of failure thresholds for each firm
	\item $\beta = n\times n$ diagonal matrix of extra failure costs for each firm.
\end{itemize}
The matrix $C$ describes the linear cross-holding relationships between firms. If a firm $i$'s market value (defined next) falls below its threshold $\theta_i$, it incurs an extra failure cost $\beta_{ii}$. We assume $C$ is column sub-stochastic as otherwise $\hat C^{-1}$ is not well-defined. Notice that this also means that $I-C$ is invertible because the spectral radius $\rho(C)<1$.

\begin{figure}
	\centering
	\includegraphics[width=0.7\textwidth]{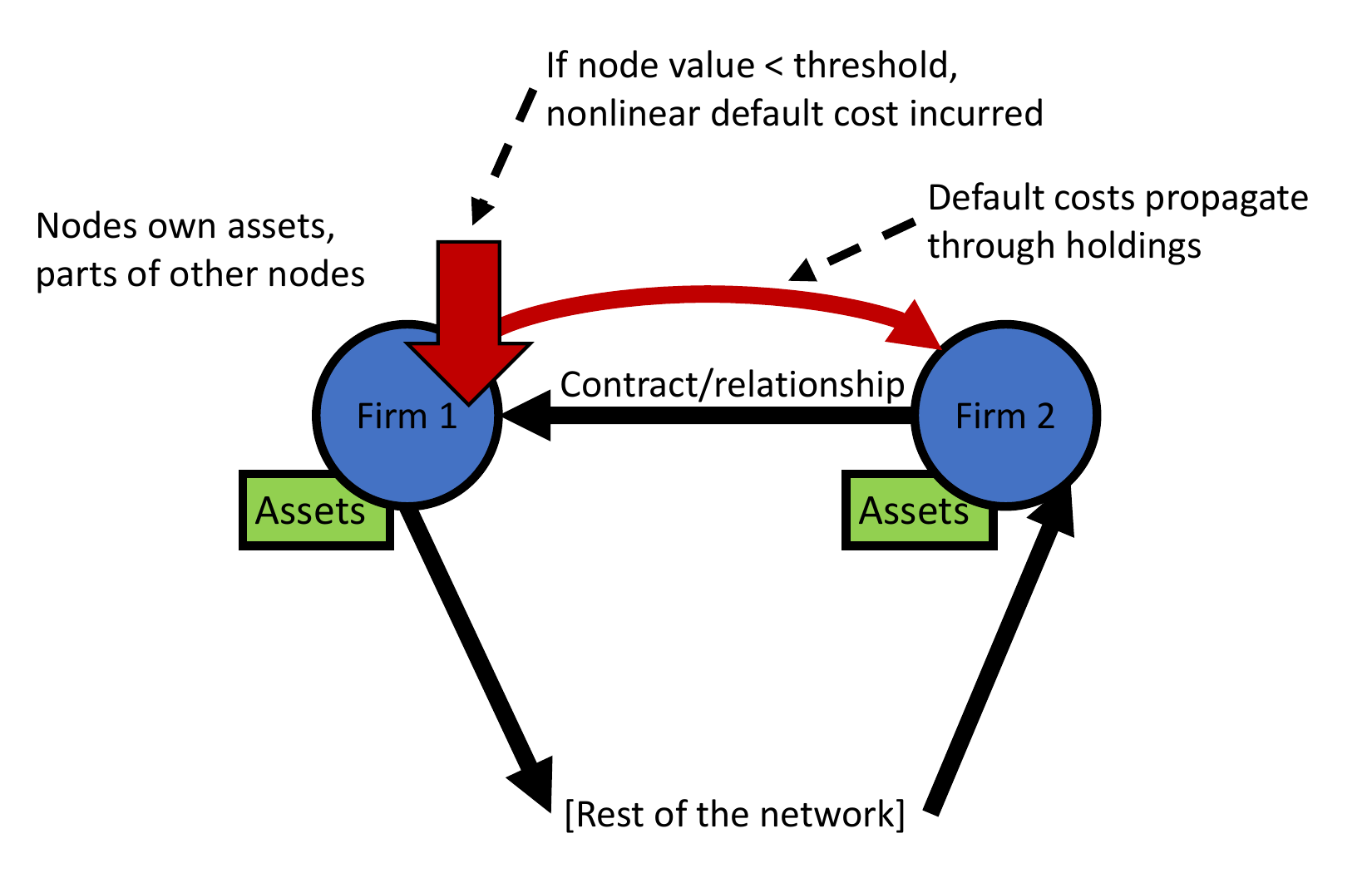}
	\caption{Financial network propagation mechanism.}\label{fig:model_diagram}
\end{figure}

The network propagates asset values and defaults across firms in the network. We illustrate this conceptually in Figure~\ref{fig:model_diagram}. $D$ describes the mapping of underlying assets (blue nodes) to firms (orange nodes). $C$ describes cross-holdings between firms. The breach of a threshold triggers failure costs, which propagate to other firms through $C$.

Firm book values are given by
$$\mathbf V = C \mathbf V + D \mathbf p - \beta \Ind_{\{\mathbf v < \bm \theta\}},$$
where $\Ind_S$ is the 1-0 valued vector indicating the entries of set $S$.
$\mathbf V$ represents the vector of all book values across the network. 
The first term $C \mathbf V$ gives the firm cross holdings, i.e., the book value of each firm contains a fraction of the values of all other book values.
The second term $D \mathbf p$ represents the value of the external asset holdings, in vector form. Finally, the last term represents bankruptcy costs, which occur in the case that the market value of the firm drops below a failure threshold.

Notice that book values inflate the value of underlying assets because asset values are counted multiple times across firms (consequently, $\|\mathbf V\|_1 \geq \| \mathbf p\|_1$ and can be arbitrarily large). A more useful measure of value is a scaling of book values by $\hat C$, accounting for the ownership share that each firm retains in itself. These are market values, which are given by
$$\mathbf v = \hat C \mathbf V = \hat C(I-C)^{-1}(D\mathbf p - \beta \Ind_{\{\mathbf v < \bm\theta\}}).$$

In \cite{jackson14}, authors show that the matrix $\hat C (I-C)^{-1}$ is column-stochastic.

\paragraph{Lattice of solutions.}
As defined, there is always a solution for $\mathbf v$. The set of solutions forms a complete lattice via Tarski's fixed point theorem. Further, supremum and infimum exist (best and worst cases). The analysis in \cite{jackson14} focuses on the best case solution as other solutions in the lattice are due to self-fulfilling failures.

\paragraph{Intervention lowers thresholds.}
Beyond the core model from \cite{jackson14}, we add a vector of intervention payments $\bm \gamma \geq 0$, which affect the default status of firms. Given an intervention profile $\bm\gamma$, firm $i$ now defaults if
$$V_i + \gamma_i < [\hat C^{-1}\bm\theta]_i.$$
This leads to post-intervention market values
$$\mathbf{\tilde v} = \hat C(I-C)^{-1} (D\mathbf p - \beta \Ind_{\mathbf V + \bm\gamma < \hat C^{-1}\bm\theta}).$$

An intervention via this mechanism effectively lowers the failure threshold of firms. This is consistent with real-world intervention mechanisms as discussed in the introduction.

\subsection{Optimal intervention}\label{sec:opt_int}

Defining an optimal intervention requires a performance measure for the severity of a cascade. In economic networks, nodes can vary vastly in size with larger and more connected institutions being more systemically important than smaller less connected institutions. A good performance measure for economic networks will be akin to total value of surviving nodes in the network. In an optimal intervention, we should seek to maximize this or, equivalently, minimize the value destroyed in a default cascade. We require an appropriate \textit{weight function} $w(S)$ that outputs the importance measure of node set $S\subseteq U$. A few obvious and simple choices for $w$ are consistent with maximizing value or minimizing value destroyed. For example: fixed node weightings of current market values of nodes, e.g.,  $$w(S) = \sum_{i \in S} v_i $$ or, relatedly, the level of failure costs $\beta$ associated with each node. In particular, these choices allow us to capture the size importance of nodes., e.g. $w(S) = \sum_{i \in S} \beta_i $. A well-defined intervention objective is then to maximize $w(S)$ where $S$ is the set of non-defaulting nodes.

A well-defined optimal intervention also requires a resource constraint. We define $b$ to be the intervention budget. Then a well-defined optimal intervention is the solution to the following optimization problem.
\begin{equation}
\label{eq:def_opt_int}
\begin{aligned}
    \max_{\bm \gamma \geq 0} \hspace{1cm} & w(S) \\
    \text{s.t.} \hspace{1cm} & \mathbf{1}^T \bm \gamma \leq b
\end{aligned}
\end{equation}
where $\mathbf{1}$ is the all-ones vector.

Toward solving this, it will be convenient to transform the problem and introduce some additional notation. We can reinterpret the intervention problem in the economic network model as the following: given an impending default cascade, how do we find an optimal intervention to optimally reduce defaults?

Suppose that the set of nodes that would default without intervention is $T$. Now reduce the system to only look at effects on the nodes in $T$, while preserving the entire network structure. 
To do this, suppose that the set of nodes that would default  without intervention is $T$. In particular, define the following
\begin{itemize}
	\item $I_T = $ diagonal matrix with $I_{uu} = 1$ for $u\in T$ and $0$ otherwise.
	
	\item $\Psi(T)$ maps to a system on the non-zero diagonal coordinates of $I_T$. Essentially, $\Psi(T)$ is the $|T|\times |U|$ matrix obtained by dropping zero rows of $I_T$.
	
\end{itemize}

We can apply the above map to transform the system to look at
$$\mathbf{\bar v} := \Psi(T) \hat C(I-C)^{-1}(D\mathbf p - \beta \Ind_{\mathbf v < \bm\theta}).$$
This transformation removes firms that don't fail without intervention, while preserving the networked connections through such nodes.
The idea is that among the firms who would fail without intervention, some of them will be saved by direct intervention. Their value would then go above the failure threshold and in particular the failure costs are reversed. In a reverse causal relation of failure,  other firms would be indirectly saved because their value would also increase.
To simplify notation, we will proceed where applicable without the bar and $\Psi$ notation, but assuming we are working in the transformed problem that only directly considers nodes $T$ on which we may intervene.
In the sequel, it is understood that the set   $T$ is fixed.

\paragraph{Intervention impact function.}
We next let the set function $f$ define the intervention impact vector on the nodes in $T$ from an intervention that reverses the failures of nodes in the set $S\subseteq T$. In particular, $f_u(S)$ is the impact on node $u$ from the intervention on nodes in $S$. As we will explain, this is given by
\begin{equation}\label{eqn:influence_fn}
	f(S) = (I-C)^{-1} \beta \Ind_S - \sum_{u\in S} I_u (I-C)^{-1} \beta \Ind_u.
\end{equation}
This accounts for the effect on book values across the network of reversing faliure costs in the $S$ nodes (the first term), which pushes other nodes closer to their failure reversal thresholds. Note that the reversal of a node's default has an effect on itself through cross-holdings. Further, notice that the intervention $\bm \gamma$ does not need to cover the cost of $\beta$ as the intervention stops this cost from being realized in market values. The second term in $f(S)$ removes this self-influencing effect from the impact function as it is instead represented in reduced intervention thresholds. Notice that $f(\emptyset) = 0$ so that the impact function is normalized.

\paragraph{Intervention thresholds.}
For the initial defaulting set $T$ and a node $u\in T$, we define the intervention threshold $\tilde\theta_u$ to describe how much book values would need to change in order for the failure of $u$ to be reversed. With some simple algebra, this is given by
\begin{equation}\label{eqn:tilde_theta}
	\tilde\theta_u = \Big[ \hat C^{-1}\bm\theta - (I-C)^{-1}(D\mathbf p - \beta \Ind_{T\setminus \{u\}}) \Big]_u
\end{equation}
This can be interpreted as the slack below threshold in the economic network. We can obtain this by taking $[\hat C^{-1}\bm\theta - (I-C)^{-1}(D\mathbf p - \beta \Ind_T)]_u$, the divergence of book value from the failure threshold (measured in book value), and subtracting the self-influencing effect described above.

\paragraph{Evaluating an intervention.}
The intervention $\gamma$ reverses the defaults of a ``seed'' set of nodes $S_0 \subseteq T$. The set $S_0$ is composed of nodes $u$ for which $\gamma_u \geq \tilde\theta_u$. We can iteratively construct subsequent sets of nodes $S_i\subseteq T$ (for $i\geq 1$) whose defaults are reversed by propagating the effects of $S_{i-1}$. This is done by adding to $S_{i-1}$ the nodes $u$ such that
$$f_u(S_{i-1}) + \gamma_u \geq \tilde\theta_u.$$

Note that the amounts in $\gamma$ can be a fraction of the thresholds of the nodes. This allows more efficient use of the budget $b$. In particular, this takes advantage of the fact that we don't have to spend as much to impact a node that already has partial impact exerted from other impacted nodes.

This leads to an optimization problem equivalent to (\ref{eq:def_opt_int}). The only change is in the restriction of $S$ to the superset $T$, which leads to the change of a constant term in the objective involving the weights of the nodes not in $T$.

\paragraph{Randomized thresholds.}
We will further consider a modified form of the problem with randomized thresholds. For instance, this is the case if there is some inherent uncertainty about what the thresholds are. In this case, a well-defined intervention problem is to optimize the expected performance measure of the intervention:
\begin{equation}\label{eq:sigma1}
    \sigma(\bm \gamma) := \EX[ w(S) | \bm \gamma],
\end{equation}
where the expectation is taken over the random thresholds. This corresponds to the \textit{fractional} intervention problem, where we select arbitrary $\bm \gamma$. In the proofs, it will be helpful to start with a simplified \textit{integral} intervention problem, where we select nodes to bail out in the initial seed set $S_0$, intervening with the full intervention threshold value. In this case, the appropriate expected performance measure of the intervention is
\begin{equation}\label{eq:sigma2}
    \sigma(S_0) := \EX[ w(S) | S_0].
\end{equation}

These optimization problems bears striking similarity to influence maximization models in social networks, like in \cite{kempe03,demaine14}, with several key differences in the forms of $f$ and $\bm\tilde\theta$, particularly in accounting for the effects of a firm on itself,  and appropriate weight functions $w$.

\section{Analytical Results}\label{sec:analytical_results}

In the previous section, we set up an economic network model and a well-defined optimal intervention problem that relates to influence maximization problems. 
We discuss these influence models in Appendix~\ref{sec:influence_max}. In the remainder of the paper, we will develop this connection, which allows us to transfer powerful tools from the influence maximization literature to the world of economic interventions.

Our proofs, given in Appendix \ref{sec:Proofs} rely on strategies used in some simpler influence maximization-like problems, such as the linear influence model. New challenges arise when reducing from the independent set problem to the economic network intervention setting, which is a class of instances of more general influence maximization-like problems.

We  prove theoretical properties of the intervention model. We prove that it is NP-hard to optimize the economic network intervention and cannot be approximated to a constant factor in polynomial time. Additionally, we prove that randomizing thresholds under appropriate assumptions yields objective functions that are monotone and sub-modular. Namely, we show that  the economic network intervention problem, when modified to consider expected values under random thresholds, is monotone submodular. Consequently, one can use the results from \cite{mossel07,demaine14} to provide an $(1-1/e-\epsilon)$-approximation in polynomial time.

\subsection{Hardness of optimal intervention}

In our first result, we show that the optimal economic network intervention problem is NP-hard. Note that this result is not a consequence of influence maximization hardness results in, e.g., \cite{kempe03,demaine14,gunnec19}. While we can transform the economic network intervention model into a form that resembles influence maximization, that does not mean that the general hardness of influence maximization extends to this case.

\begin{theorem}\label{result:interventions_hard}
	Let $(C,D,\beta, \bm\theta, \mathbf p)$ be a financial system with $n$ firms and deterministic thresholds $\theta$, and let $0 \leq \ell < \alpha \leq 1$. Suppose $\alpha n$ firms fail in the financial system equilibrium. Then it is NP-hard to determine whether there exists an intervention $\gamma_i\geq 0$ with $\| \bm\gamma \|_1 \leq b$ such that at most $\ell n$ nodes fail after the intervention.
\end{theorem}

\begin{center} \hyperlink{pf:interventions_hard}{\texttt{[Link to Proof]}} \end{center}

Recall that the meaning of NP-hard is that a general instance of the problem is hard, while naturally there may be parameter values (e.g., budget of zero) for which the problem may not be hard. We prove this for $w$ with equal weights, which means that the problem is NP-hard for general weighting functions. The proof is a reduction from independent set.

A consequence of the theorem is the following corollary describing hardness of approximation.

\begin{corollary}\label{result:approx_hard}
	Optimal economic network intervention cannot be approximated to within a constant factor in polynomial time.
\end{corollary}

Additionally, note that it may be much harder to approximate the optimal intervention problem than proven in Corollary~\ref{result:approx_hard}. For example, similar influence maximization problems have approximation difficulties that scale in the dimensions of the system \cite{demaine14,gunnec19}.

\begin{remark}
(Default hierarchies) While we can identify the hierarchy of defaults in the initial cascade, which the intervention aims to counteract, this does not make the optimal intervention problem, in general, easier. Consistent with Corollary~\ref{result:approx_hard}, simply intervening in a layer of this hierarchy, which would prevent all defaults in subsequent layers, does not guarantee a good approximation to optimal intervention. In particular, only intervening across an entire layer may be far from optimal if all layers are very wide. This is the case in the 2008 financial crisis but closer to the case in the 2020 Covid crisis, when much of the economy was shut down. Intuitively, the initial default hierarchy doesn't describe all possible sequences of default; making some intervention payments in turn alters the effective hierarchy sequence. Similar ``activation hierarchies'' are also present in the influence maximization literature and do not make those problems easier either. The default hierarchy does not help us devise an approximation algorithm in general, which remains true when  using different objective weighting functions, including current total market cap of solvent firms.
\end{remark}

\subsection{Approximation with randomized thresholds}

We now establish that a modified form of the optimal intervention problem can be well-approximated in polynomial time. The modification incorporates randomized thresholds and reframes the problem to optimize \emph{in expectation}. For instance, this can be done by treating thresholds as random variables uniformly distributed over any given uncertainty range. This can be done more generally with different threshold distributions, as we will discuss. In essence, the combinatorial complexity problems disappear in expectation.\footnote{Since the range of the random variables can be arbitrarily small, this is like saying that the approximation problem is difficult only on measure 0 sets.}

We first show that the intervention problem with random thresholds is monotone submodular, connecting with results from \cite{mossel07} and \cite{demaine14}. As a result, a greedy hill-climbing algorithm provides a $(1-1/e-\epsilon)$-approximation using results from \cite{nemhauser77,nemhauser78}.

Our next result establishes that the intervention impact function in the intervention problem is monotone submodular.

\begin{proposition}\label{result:f_submodular}
	The function $f$ from Eq.~\ref{eqn:influence_fn} is monotone increasing and submodular.
\end{proposition}

\begin{center} \hyperlink{pf:f_submodular}{\texttt{[Link to Proof]}} \end{center}

We need a few assumptions to prove that the objective $\sigma$ for intervention problem under random thresholds is monotone submodular. The first assumption describes the randomization of thresholds and is necessary for the results of \cite{mossel07} to apply. It allows very general distributions of thresholds, an example of which is uniform distributions.

\begin{assumption}
	For $u \in U$, random thresholds $\theta_u$ are independent with distribution function $F_u$ such that $F_u \circ f_u$ is monotone increasing submodular.
\end{assumption}

The next assumption is that the intervention impact function $f$ is normalized--all nodes in $T$ start out in default. With fixed thresholds, this is a property of $f$, as noted in the previous section. If we make thresholds $\bm\theta$ random in the economic network setting, this is more complicated because the corresponding intervention thresholds $\bm{\tilde\theta}$ in \eqref{eqn:tilde_theta} could be zero or negative depending on the realization of thresholds, and, if this occurs, the resulting $\bm{\tilde\theta}$ distributions are not independent. This can be solved in two ways that keep the initially defaulting nodes technically fixed: (1) the randomization in thresholds can be associated with $\bm{\tilde\theta} \geq 0$ instead of with $\bm\theta$, or (2) the problem can be reformulated: $\bm{\tilde\theta}$ becomes the positive part in \eqref{eqn:tilde_theta}, initial defaults are fixed, $f(\emptyset):=0$, and when $\tilde\theta_u=0$, $u$ can be added to the seed set with cost $0$ (and so will be added first).

\begin{assumption}
	The intervention impact function $f$ is normalized, i.e., $f(\emptyset) = 0$.
\end{assumption}

The final assumption concerns the function describing node weighting in the objective. The weight function describes how valuable it is to reverse the defaults of a given set of nodes.

\begin{assumption}
	The weight function $w: 2^{U} \rightarrow \mathcal{R}_+$ is normalized, monotone, and submodular.
	\label{as:weightfunction}
\end{assumption}

A very flexible range of functions satisfies this assumption. For example, the cardinality function, which weights each node equally  would be interpreted as minimizing the number of defaults. As discussed in the previous section, for economic networks, we generally want to incorporate the size and importance of nodes into this function, as we want to maximize something like the total welfare of surviving nodes in the network or minimize the value destroyed in a default cascade. Any fixed weighting of nodes also obeys this assumption, including weighting by the current market values of firms or, relatedly, the level of failure costs associated with each node.

Under these assumptions, the intervention objective function--e.g., the expected number of defaults under a given intervention--is monotone submodular based on results from \cite{mossel07}, as formalized in the next result.

\begin{theorem}\label{result:intervention_submodular}
	Given assumptions 1-3 and an instance of the economic network intervention problem with random thresholds, then $\sigma(S_0)$ and $\sigma(\bm \gamma)$ are normalized, monotone, and submodular.
\end{theorem}

\begin{center} \hyperlink{pf:intervention_submodular}{\texttt{[Link to Proof]}} \end{center}

Then following the application of results in \cite{kempe03}, there is a greedy $(1-1/e - 1/\text{poly}(n))$-approximation algorithm for optimizing the expectation, as formalized in the next corollary. The integral and fractional forms of this greedy algorithm are described in Appendix~\ref{sec:algorithms} (Algorithm~\ref{alg:int_greedy} and Algorithm~\ref{alg:frac_greedy}).

\begin{corollary}\label{result:greedy_approx}
	Given assumptions 1-3, there exists a polynomial-time greedy $(1-1/e-\epsilon)$-approximation for maximizing $\sigma(S_0)$ and $\sigma(\bm \gamma)$ subject to budget $b$.
\end{corollary}

\begin{center} \hyperlink{pf:greedy_approx}{\texttt{[Link to Proof]}} \end{center}

\subsection{Identifying large failure cascade scenarios}\label{subsec:large_cascades}

We now show how these results translate into related economic network problems. We start by showing that it is also NP-hard to identify the worst case failure scenarios given a maximum sized aggregate shock to asset values $\mathbf p$. Like the intervention problem, there is a $(1-1/e-\epsilon)$-approximation under random thresholds. While we may not generally be interested in uncovering the strictly \textit{worst} failure scenarios, we will see that the results about this \textit{do} lead to very useful and interesting applications regarding sampling tail events in general.

\begin{theorem}\label{result:max_shock_hard}
	Suppose $(C,D,\beta,\bm\theta,\mathbf {p_0})$ is an instance of an economic network and asset prices evolve to $\mathbf {p_1}$ such that $\|\mathbf {p_0}\|_1 - \|\mathbf {p_1}\|_1 \leq b$ for some maximum aggregate shock $b>0$. Let $0<\ell<1$. Then it is NP-hard to determine if a failure cascade of size $\ell |U|$ is possible in $(C,D,\beta,\bm\theta,\mathbf p_1)$.
\end{theorem}

\begin{center} \hyperlink{pf:max_shock_hard}{\texttt{[Link to Proof]}} \end{center}

The reduction from independent set again implies a corollary result that the optimum is hard to approximate up to a constant factor in polynomial time. As in the intervention case, when reframed in terms of expectations under random thresholds, a greedy $(1-1/e-\epsilon)$-approximation again applies.

As alluded above, we now develop two useful and interesting consequences of these results: (1) it is computationally hard to calculate expected values of nodes in the economic network, and (2) approximation methods can be applied for importance sampling to identify instances that lead to events in the tail. The depth of sampling in the tail can be tailored by choosing the parameter $b$. This can be very valuable for the application of stress testing.

\paragraph{Hardness of calculating expected values.}
We next demonstrate a consequence of Theorem~\ref{result:max_shock_hard}, namely it can be computationally hard to calculate expected values of firms in an economic network even if we have perfect information about the underlying setup. Consider a simple setting in which the prices of underlying assets $\mathbf p$ are i.i.d. Bernoulli distributed 0-1 with probability $q$. The probability that a specific set of $b$ assets fail is $(1-q)^b$, which is non-vanishing in the scale of the network and so non-negligible for the calculation of expected value of firms when the problem is large (and potentially computationally complex). Since it is NP-hard to determine whether a large failure cascade can occur with that probability, it is in turn NP-hard to determine if the expected value is above some given level. Further, the ability to approximate will depend on the failure costs $\beta$ in the network, which could be arbitrarily large in the general case, suggesting that approximation is also difficult in general under fixed thresholds.

This compares to what is typically done in financial models in practice. Firms are typically treated in isolation, i.e., not part of a network model. In this case, firm defaults are treated as independent or perhaps correlated through a simple copula. Such distributions of credit risk, such as produced by a Gaussian copula, fail to capture clustering of defaults. The resulting probability that a given fraction of firms default is exponentially unlikely as the number of firms grows, and so this computational problem does not arise in those simple models. Naturally, the assumption that firm defaults are independent is flawed, and so the complexity problems that we describe in calculating expected values  apply in realistic settings.

\paragraph{Importance sampling of tail events.}
While it is NP-hard to identify the worst case failure scenarios given a maximum sized aggregate shock $b$ in an economic network, it is possible to identify scenarios that approximate this up to a $(1-1/e-\epsilon)$ factor with random thresholds. As a result, we can apply influence maximization approximation methods to identify instances of shocks that lead to events in the tail of similar size to the parameter $b$ that is chosen (or indeed for a variety of $b$ chosen). A common task in finance is to stress test a financial system subject to aggregate shocks up to a particular size. Direct Monte Carlo approaches will tend to underestimate risks because random samples are unlikely to contain many of the extreme default scenarios, especially in a large multi-dimensional space. Importance sampling using this new suite of approximation algorithms thus unlocks a valuable new way to sample tail events where it was otherwise difficult.

\section{Application to WIOD dataset}\label{sec:simulation}

To demonstrate the use of our results, we consider an application of influence maximization algorithms to an economic network. We construct instances of the economic network intervention problem based on the World Input Output Database (WIOD).  The data is openly available at \url{http://www.wiod.org/home}. We simulate a number of possible shocks to the resulting network and demonstrate that by adapting influence maximization algorithms, we can derive effective interventions using relatively modest budgets. As we might expect, we see decreasing returns to scale in the size of the budget. 

The simulations we perform are intended as a proof of concept of a realistic-looking setup based on real underlying data. We stress that many parts of the setup for which data is not available remain stylized: in particular underlying assets, thresholds, failure costs, and distribution of shocks to underlying asset values. Additionally, there is naturally uncertainty about economic network structure as described by the dataset and aggregation effects from grouping entire industries of firms into single nodes.

\begin{figure}
	\centering
	\includegraphics[width=\textwidth]{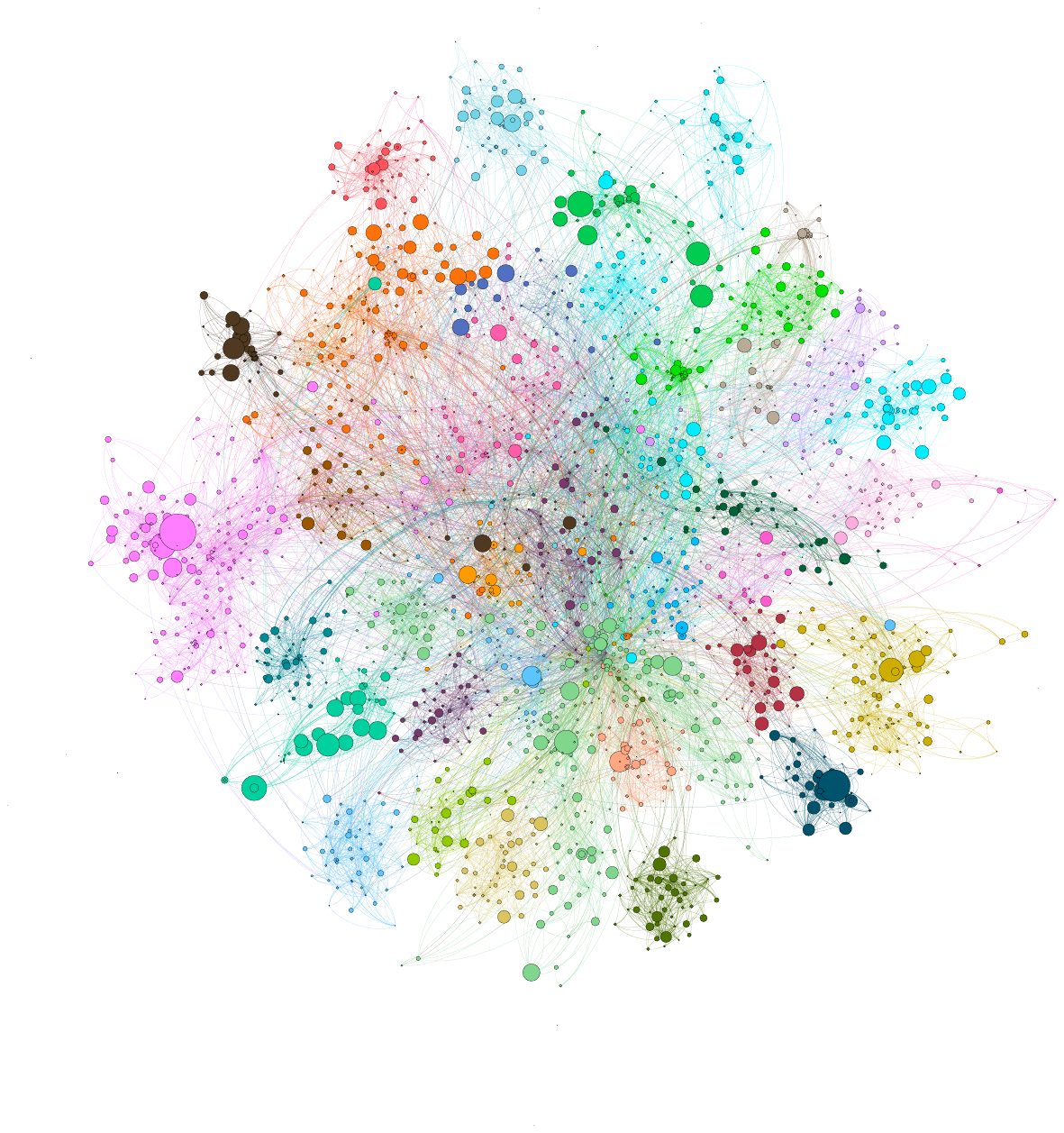}
	\caption{Economic network structure inferred from World Input Output Database (WIOD)}\label{fig:network}
\end{figure}

Our code for intervention approximation algorithms and simulation implementation is openly available at \url{https://github.com/aklamun/optimal_intervention}.
The network visualization of the data is provided in Figure \ref{fig:network}.

\subsection{Simulation setup}

The WIOD dataset (see, e.g., \cite{timmer15}) describes the flow of resources in dollar value between different economic sectors within different nations (intermediate demand) and national final demand (e.g., GDP components, such as consumption, investment, government expenditure). The dataset includes this information for 2464 distinct economic sectors spread between 28 EU countries and 15 other major countries for the years 2000-2014.

We construct an economic network from the 2014 dataset in the following way:
\begin{enumerate}
	\item We set the number of nodes to $n$, which represents the number of columns in the dataset that refer to economic sectors or final demand components;
	\item We set up an $n\times n$ array of flows between nodes from dataset, with zero rows for final demand components;
	\item We transpose components of any negative entries in the array;
	\item We scale columns to sum to 1 (inclusive of {\bf value added}, a row in the dataset that is not included in the array) or 0 if a zero column; Value added is traced by all labor and capital that is directly and indirectly needed for the production of final manufacturing goods, see \cite{timmer15};
	\item We set diagonals in the array to zero to obtain $C$;
	\item We fix unnecessarily bad conditioning in $C$ by removing nodes with near zero {\bf value added} (columns referring to households);
	\item We set the vector $D\mathbf p$ to equal the output of each node at basic prices (this is the \texttt{TOT\_GO} row in the dataset);
	\item We set the vector $\bm\theta = \hat C (I-C)^{-1} D\mathbf p - \text{\bf value added}$, which gives the market value assuming no defaults from which we subtract value added;
	\item We let the diagonal matrix $\beta$ with diagonal entries $0.1\cdot \text{\bf value added}$.
\end{enumerate}

The vector $D\mathbf p$ above represents initial asset values. We sample shocks to these asset values by sampling a shock vector $r$ such that the shocked asset prices are given by the component-wise multiplication $D \mathbf p\cdot (1+r)$. The shock vector $r$ is sampled from a $m$-dimensional normal distribution with the following specifications intended to sample a range of large deviations:
\begin{itemize}
	\item Common correlation factor $\rho = 0.6$,
	\item Marginal distributions have $\sigma = 0.15$ and drift $a=-0.3$,
	\item Shocks bounded by $0$ such that $1+r_i = \max (1+r_i, 0)$.
\end{itemize}
Recall that $D\mathbf p$ are underlying asset prices, and market values will have additional inter-relation and correlation from the network process.

\subsection{Intervention algorithms}

Based on our main results in the previous section, under appropriate assumptions and randomization of thresholds, the network intervention problem is monotone submodular. 
In this case there are known greedy algorithms that provide $(1-1/e-\epsilon)$-approximations. For the reader's convenience, we provide these explicitly in Appendix~\ref{sec:algorithms} (Algorithm~\ref{alg:int_greedy} and Algorithm~\ref{alg:frac_greedy}). The general structure of these greedy algorithms is to start with an empty seed set $S_0$ and, iteratively, add the node $u$ to $S_0$ that gives the maximum marginal gain. Since the thresholds are random, determining the maximum marginal gain in each step involves estimating the expected size of resulting cascades $\sigma(S_0\cup\{u\})$ for a number of nodes $u$. This is typically done through Monte Carlo estimation of the expectation integral. For large networks, for which these integrals are very high-dimensional, the Monte Carlo approximations become prohibitively slow, although still within polynomial time with the Monte Carlo capped at a constant factor.\footnote{As an area of future research, it would be interesting to examine whether asymptotic results on  the size of the cascade \'a la \cite{amini2016resilience} could replace part of the Monte Carlo approximations.}
This is the case for the size of networks in these simulations.

In practice, heuristic algorithms are used in influence maximization to try to estimate the greedy algorithm in faster time with large success. For instance \texttt{DiscountFrac} used in \cite{demaine14} starts with an empty seed set $S_0$ and iteratively adds the node $u$ to $S_0$ that would exert the most total impact on the remaining defaulting nodes. In particular, given the initial intervention seed set $S$ at the beginning of a step, \texttt{DiscountFrac} picks the node $u$ that maximizes $\|f(\{u\})\Ind_{A\setminus\{u\}}\|_1$ for remaining uninfluenced set $A$. We provide an explicit description of \texttt{DiscountFrac} in Appendix~\ref{sec:algorithms} (Algorithm~\ref{alg:discount_frac}).

In our simulations, we adapt \texttt{DiscountFrac} to choose the node $u$ that maximizes
$$\frac{\|f(\{u\})\Ind_{A\setminus\{u\}}\|_1}{\tilde\theta_u - f_u(S)},$$
where $S$ is the currently influenced set. This accounts for the cost to influence node $u$ in the current step, given that economic network thresholds can vary significantly in size. For full implementation details of this adaptation, we refer to our public code repository at \url{https://github.com/aklamun/optimal_intervention}. The heuristic algorithm is conceptually very similar to the ideal fractional greedy algorithm. Although it does not come with the same theoretical approximation guarantees, it  performs well in practice.

\subsection{Simulated interventions}

We simulate 5000 shocks and apply the adaptation of \texttt{DiscountFrac} to approximate the resulting optimal intervention problems. In this setting, we explore the effectiveness of a range of targeted intervention sizes.

%

\begin{figure}
	\centering
	\begin{subfigure}{0.5\textwidth}
		\includegraphics[width=\textwidth]{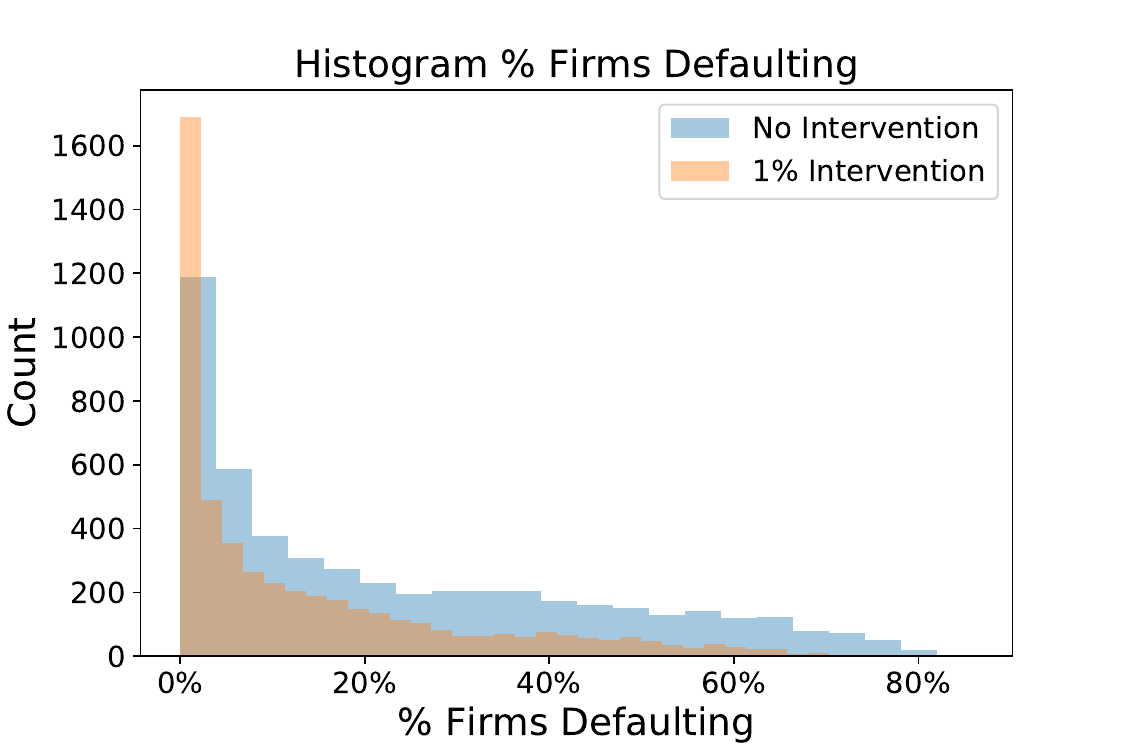}
		\caption{}\label{subfig:1d_hists}
	\end{subfigure}
	\begin{subfigure}{0.43\textwidth}
		\includegraphics[width=\textwidth]{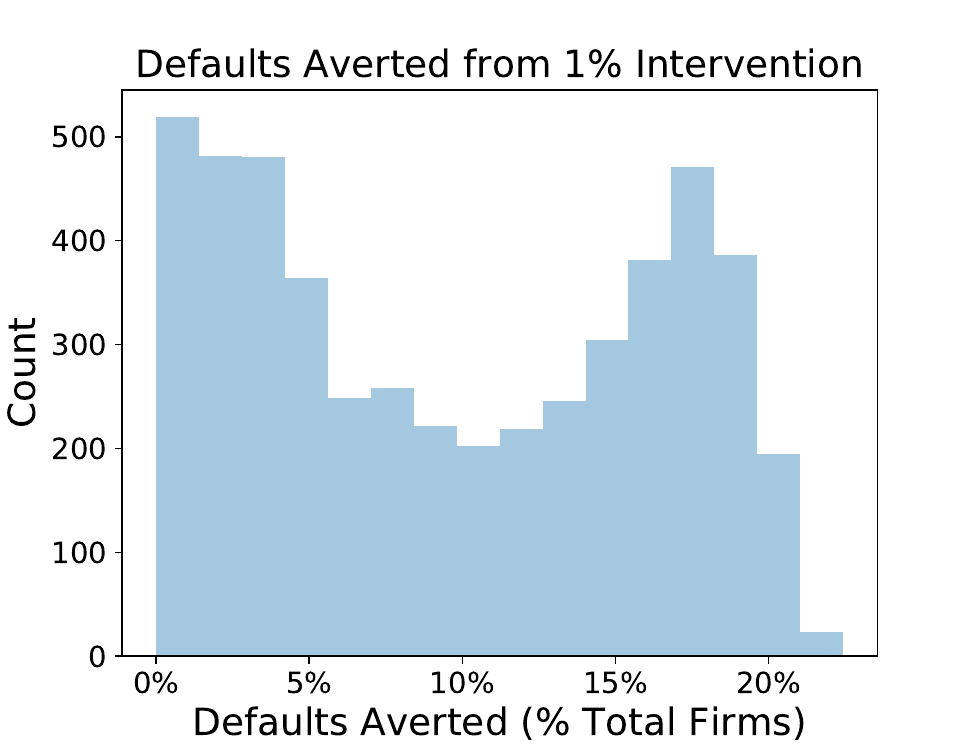}
		\caption{}\label{subfig:1d_hists_diff}
	\end{subfigure}
	\caption{Histogram densities of defaults under 1\% asset value intervention and no intervention.}\label{fig:1d_hists}
\end{figure}

Figure~\ref{fig:1d_hists} depicts  the percentage of firms defaulting under certain intervention  scenarios. In particular, we  compare the effects of a 1\% targeted intervention to no intervention. Figure~\ref{subfig:1d_hists} shows histogram densities of firm defaults under the sampled shocks, illustrating that the 1\% intervention effectively reduces the tails of this distribution. 

Figure~\ref{subfig:1d_hists_diff} shows histogram densities of defaults averted under the 1\% intervention relative to no intervention, which also illustrates the effectiveness. An interesting feature is the bimodal distribution of defaults averted from targeted intervention. One hypothesis to consider is that this is a result of the network cluster structure itself: there are several clusters in the network, and firms within the same cluster are more likely to default (or avert default from a nearby intervention) together.

\begin{figure}
	\centering
	\includegraphics[width=0.5\textwidth]{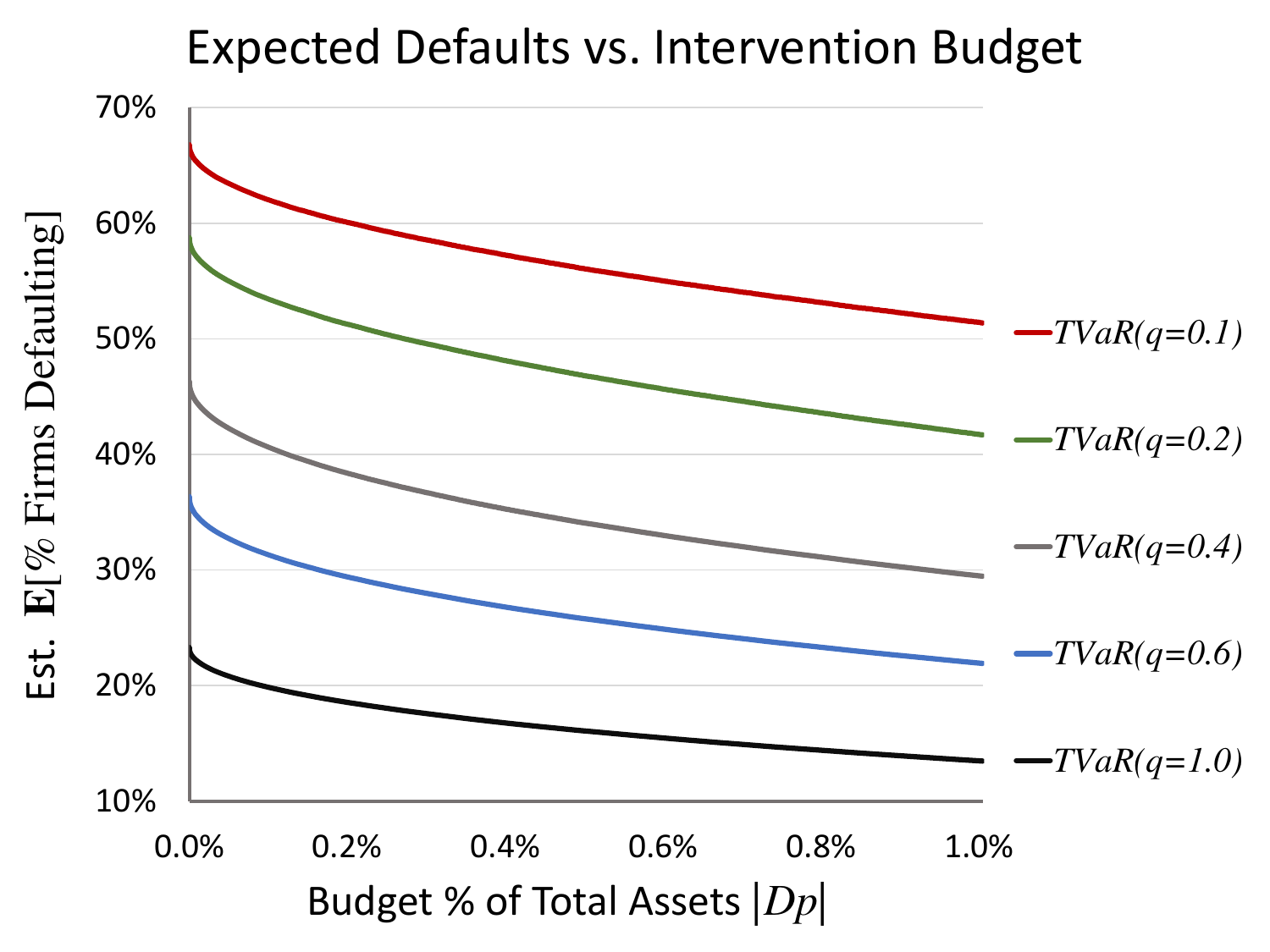}
	\caption{Simulation TVaRs with quantiles $q$ for a range of intervention budgets.}\label{fig:tvars}
\end{figure}

Figure~\ref{fig:tvars} depicts the experimental Tail Value at Risk (TVaR) of default cascade size for different quantiles $0< q \leq 1$. $\text{TVaR}(q)$ is a conditional expectation, conditioned on events falling in the $q$-th quantile of outcomes:
$$\text{TVaR}(q;b) = \EX \left[ \frac{|A|(b)}{|U|} \hspace{0.2cm} \Big\vert \hspace{0.2cm} |A|(0) \geq \text{VaR}\big(|A|(0);q\big) \right],$$
where $|A|(b)$ outputs the number of defaulting firms given budget $b$, $|U|$ is the number of total firms, and $\text{VaR}(X;q)$ is the $q$-quantile of random variable $X$.
Note that in our case $q$ is a quantile of a distribution that is already modeling negative outcomes in these simulations. Also recall that $q=1$ gives the unconditional expectation.

Figure~\ref{fig:tvars} demonstrates that relatively small budgets effectively reduce systemic risks as measured by TVaR. Experimental numbers for the percentage reduction in TVaR using an intervention budget of 1\% of initial assets is presented in Table~\ref{table:tvars}.

\begin{table}
	\begin{center}
		\begin{tabular}{c|c}
			$\mathbf{q}$	&	\textbf{\% Reduction in} $\mathbf{\text{TVaR}(q)}$ \\
			\hline
			$0.1$		&	$23\%$ \\
			$0.2$		&	$29\%$ \\
			$0.4$		&	$36\%$ \\
			$0.6$		&	$40\%$ \\
			$1.0$		&	$42\%$
		\end{tabular}
	\end{center}
	\caption{Percentage change in TVaR with quantile $q$ of default cascade size resulting from targeted intervention with budget 1\% of total initial assets.}\label{table:tvars}
\end{table}

\subsection{Efficiency of intervention}
We end this section by  exploring the efficiency of intervention.
The question of computational efficiency is clear because the problem is in general NP hard: optimizing naively would be quite daunting (and completely intractable given the even modest size of the network). A naive approach would be to to consider every subset of nodes on which to intervene.  In absence of influence maximization approximation methods, one would need to resort to  heuristics such as  (1) intervening on ``systemically important" firms first, and (2) intervening on the first layers of the default hierarchy. Neither of those heuristics have good guarantees and the size of value alone cannot be a measure of systemic importance, see e.g. \cite{ofr17} and the references therein.

Our  influence maximization method can be applied for any weight function $w(S)$ that satisfies the Assumption \ref{as:weightfunction}. Our approach is computationally efficient and we have performance guarantees. 
The fact that we can consider multiple weight functions for the same intervention algorithms allows us to examine also a notion of economic efficiency.
Using the cardinality weight function amounts to minimizing the number of defaults subject to the given budget.
We now consider the weight function represented by the sum of the market value of the nodes $$w(S) = \sum_{i \in S} v_i. $$ In this case, the goal of intervention  is to maximize value. 
In heterogeneous economic networks, we can  consider multiple objectives  in order to assess the efficiency of intervention. Since firms differ in terms of value, we expect that the additional value saved decreases with the number of saved firms. 
The approximations we provide using influence maximization methods are closer to a policy that intervenes on  "systemically important" nodes first. With this approach, the systemic importance of a node  is determined by the algorithm itself and combines the value of the firm and their position in the network.

\begin{figure}
	\centering
	\includegraphics[width=0.45\textwidth]{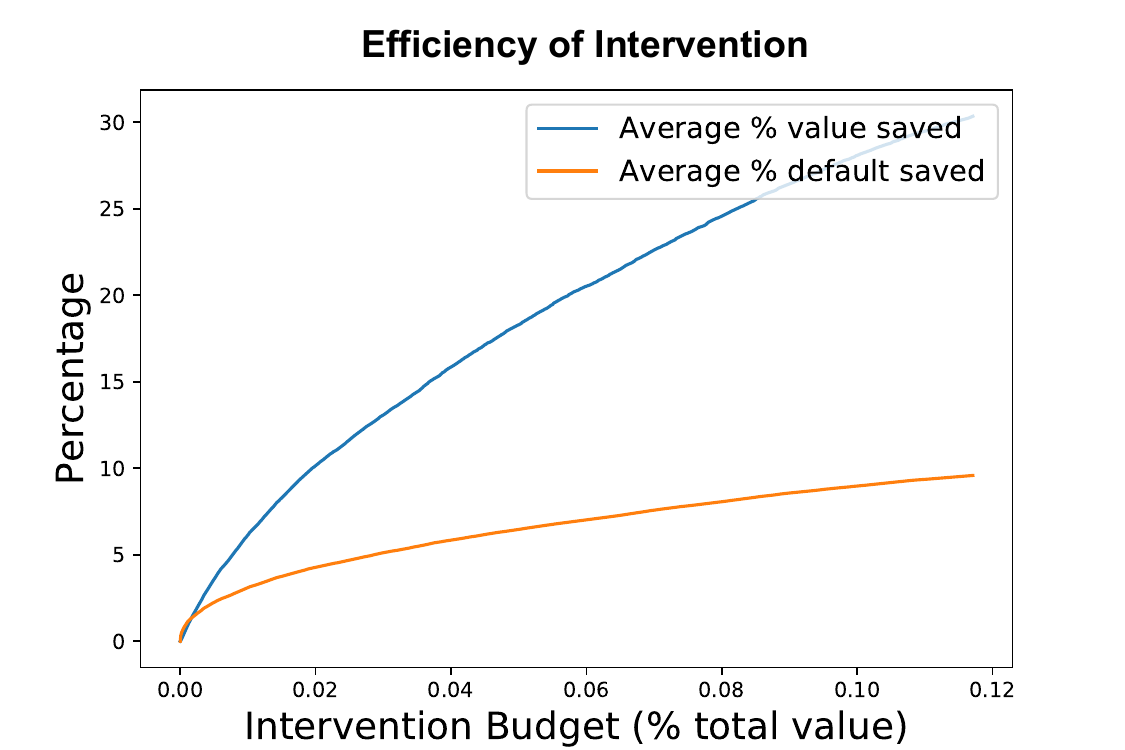}
	\caption{Efficiency of intervention}\label{fig:efficiency}
\end{figure}

In Figure \ref{fig:efficiency} we plot the percentage value  and the percentage of firms saved by intervention as a function of the intervention budget. 
These plots both demonstrate diminishing returns, although  less so when the criterion is the value saved. 
When the budget is sufficiently high, the number of firms that are being saved stays relatively flat, whereas the value saved still exhibits significant increases. This   means that the intervention set changes, and the reason why additional value is being saved is the network effects. 

\begin{figure}
	\centering
	\includegraphics[width=0.45\textwidth]{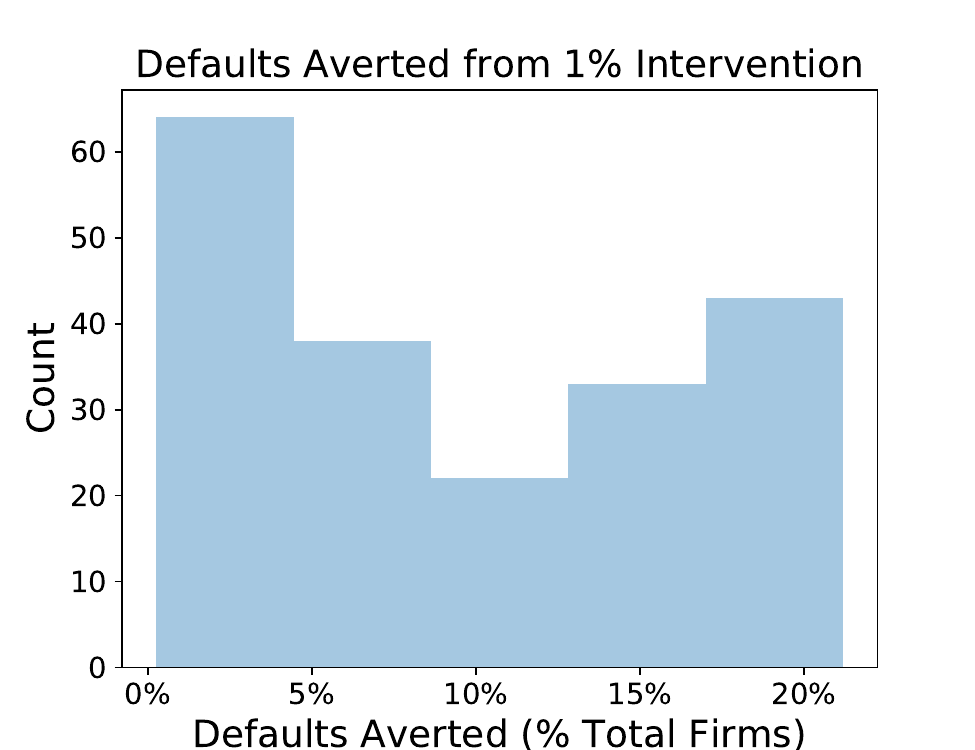}
	\includegraphics[width=0.45\textwidth]{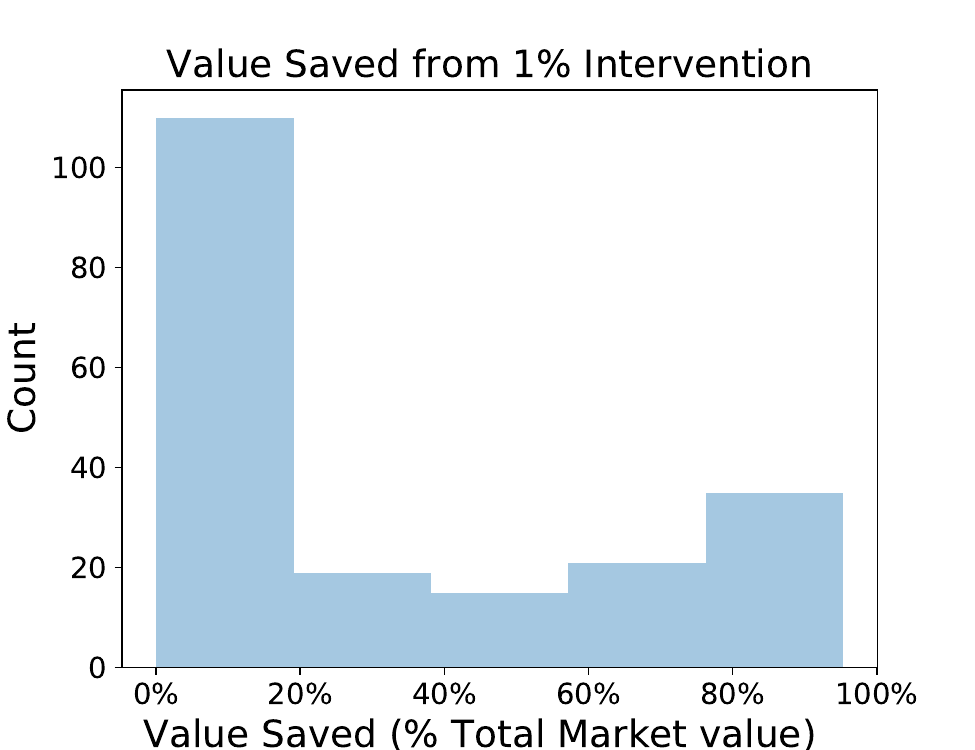}
	\caption{Defaults averted vs. value saved by  intervention}\label{fig:hist_efficiency}
\end{figure}

Next, in Figure \ref{fig:hist_efficiency} we plot the histogram of defaults averted vs. value saved across simulated shocks for a fixed budget representing $1\%$ of the total initial value. The histogram of the defaults averted is rather flat, whereas we note a more u-shaped histogram for the histogram of the value saved. This is consistent with well known phase transition  phenomena in networks: shocks either die out quickly or reach a large fraction of the network, but there are few intermediate situations. In the cases where network contagion is high, intervention proves highly effective and saves a large fraction of the network value.

\section{Conclusion}
We have shown that the optimal intervention problem is NP-hard under fixed failure thresholds.  Given a network, one essentially needs to choose a set of firms among those who would otherwise default and reverse their defaults. The choice of such firms saves the maximum value.
Other related problems are also shown to be computationally hard, even if we have perfect information about the underlying setup. In particular, given a  maximum aggregate shock, it is computationally hard to determine if there is a distribution of this shock across firms leading to a given fraction of the network to fail.
In turn, when thresholds are random, these problems allow $(1 - 1/e - \epsilon)$-approximations. Failure thresholds represent the points where shareholders of the firm decide to cease the operations and liquidate the asset.  In reality thresholds  could be based on the expectations of large cascades and large scale liquidations. Given the complexity issues in assessing which shocks lead to such extreme scenarios, it would be interesting to explore further how strategic shareholders would make their threshold choices.

Using the approximation algorithms, we  evaluate the performance of intervention under a large number of shocks. We remark a significant reduction of Tail Value at Risk of the default cascade size, even under a small intervention budget relative to total assets. This can be explained by the fact that the solution to the optimal intervention problem unveils a hierarchical or causal structure of defaults, and in practice it selects a relatively small set to directly intervene on. Most of the default cascade is then averted indirectly, by reversing failure costs and network effects.

\section{Acknowledgements}
The authors would like to thank Sid Banerjee for helpful discussions in the initial stages of this project, and to Kristina Tian for research assistance. This paper is based on work supported by  NSF CAREER award \#1653354 and a Bloomberg Fellowship.

\bibliographystyle{acm}

\appendix

\section{Overview of Influence Maximization}\label{sec:influence_max}

Our analysis builds on influence propagation research in social networks. We provide an overview of this to aid the reader. This work has historically studied processes like diffusion of technological innovation, beliefs, product adoption, and viral content. A natural question is how to engineer such a viral cascade given information about the network. 

A model for this problem is specified as follows:
\begin{itemize}
	\item $U$ is the set of nodes in the network.
	\item $f(S)$ a set function that outputs the vector of influence exerted by the activation of node set $S\subseteq U$ on each node in $U$ (i.e., $f_u(S) = $ influence exerted on node $u$). We assume $f(\emptyset) = 0$.
	\item $w(S)$ outputs an importance weighting of node set $S$. In the simplest setting, each node is weighted by 1.
	\item $\bm{\tilde\theta}$ is the vector of thresholds for each node. A node $u$ becomes \emph{activated} if the influence exerted on it is $\geq \tilde \theta_u$.
	\item $b$ is the budget for influencing nodes.
\end{itemize}

\paragraph{Integral Influence Maximization,} studied in \cite{kempe03}, focuses on maximizing the weighted number of activated nodes by finding an optimal seed set $S_0$ to activate with payments of size $\tilde \theta_u$ for each $u\in U$ subject to budget $b$. An influence cascade is calculated in stages. Given an initial set of activated nodes $S_0$, we construct the set of nodes $S_i$ (for $i\geq 1$) activated by the set $S_{i-1}$ by adding the nodes $u$ such that 
$$f_u(S_{i-1}) \geq \tilde \theta_u.$$
The cascade process converges to a final set of activated nodes $S$. The optimization problem is
$$\begin{aligned}
\max_{S_0\subseteq U} \hspace{1cm} & w(S) \\
\text{s.t.} \hspace{1cm} & \sum_{u \in S_0} \tilde\theta_u \leq b.
\end{aligned}$$

\paragraph{Fractional Influence Maximization,} studied in \cite{demaine14}, is a generalization of the integral case. In this problem, we choose a payment vector $\mathbf{x}$ subject to budget $b$ to exert influence on seed nodes. An influence cascade is again calculated in stages. An initial set of activated nodes $S_0$ is composed of nodes $u$ for which $\mathbf x_u \geq \tilde\theta_u$. We construct the subsequent sets of nodes $S_i$ (for $i\geq 1$) activated by the set $S_{i-1}$ by adding the nodes $u$ such that
$$f_u(S_{i-1}) + \mathbf x_u \geq \tilde\theta_u.$$
Note that this assumes that direct influence is additive with influence from other vertices in the network, in the sense that node activated in next stage if and only if this condition satisfied.
The cascade process converges to a final set of activated nodes $S$. The optimization problem is
$$\begin{aligned}
\max_{\mathbf x\geq 0} \hspace{1cm} & w(S) \\
\text{s.t.} \hspace{1cm} & \mathbf{1}^T \mathbf{x}\leq b
\end{aligned}$$
where $\mathbf{1}$ is the all-ones vector.
The amounts can be a fraction of the thresholds of the nodes. This allows more efficient use of budget $b$ to influence an effective seed set $S$. In particular, this takes advantage of the fact that we don't have to spend as much to influence a node that already has partial influence exerted from other influenced nodes.

For simple influence models, like the Linear Threshold Model and Triggering Set Model, these problems are NP-hard, as shown in \cite{kempe03} and \cite{demaine14}. Further, they are also hard to approximate  within any general nontrivial factor.

However, when we consider a modified problem with randomized thresholds--e.g., if activation thresholds for influence are uniform random variables--then the problem changes enough in expectation to lower complexity. In particular, the expected cascade size $\sigma(S_0) := \EX[w(S) | S_0]$ from a given seed set $S_0$ (with similar definition for $\sigma(\mathbf x)$) is monotone submodular and allows a greedy approximation that is provably within $(1-1/e)\approx 63\%$ of optimal (\cite{kempe03},\cite{demaine14}). \cite{mossel07} proved this for more general threshold models and distributions for $\tilde\theta$. In particular, letting $F_u$ be the distribution function of $\tilde\theta_u$, $\sigma(S_0)$ is monotone submodular given that the following functions are monotone submodular: $f$, $w$, and $F_u \circ f_u$ for all $u\in U$. We define these greedy algorithms explicitly in Appendix~\ref{sec:algorithms}.

In the typical influence maximization problem, a node in $S$ does not exert influence on itself. This is complicated in the economic network intervention problem because the reversal of a node's default has an effect on itself through cross-holdings. There are also differences in $\theta$ and $w$.

\section{Proofs}
\label{sec:Proofs}
\noindent\rule{\textwidth}{1pt}
\paragraph{Theorem~\ref{result:interventions_hard}} \hypertarget{pf:interventions_hard}{}
\begin{proof}
	We will reduce from the independent set problem to an instance of the economic network intervention problem.  Our reduction strategy follows \cite{gunnec19} (for the linear influence model), but we note it requires additional steps to reduce independent set to the economic network intervention setting, which is a class of instances of more general influence maximization-like problems.
	
	In the independent set problem, we are given an undirected graph $G=(U,E)$ with nodes $U$ and edges $E$. Given a number $k$, we ask if there is an independent set in $G$ of size $k$.
	
	\paragraph{Reduction gadget.}
	For the reduction, construct a bipartite graph $G' = (U_1\cup U_2, E')$ as follows:
	\begin{itemize}
		\item Add each node in $G$ to $U_1$. Attach thresholds $\frac{1}{|U|}$ to these nodes.
		\item For each edge $\{i,j\} \in E$, add a node $u$ to $U_2$ and add directed edges $(i,u),(j,u)$ to $E'$. Attach edge weights $\frac{1}{|U|}$ and thresholds $\frac{1}{|U|}$.
		\item For each possible pair $\{i,j\} \notin E$, add two nodes $u,w$ to $U_2$ and add directed edges $(i,u),(j,w)$ to $E'$. Attach intervention weights $\frac{1}{|U|}$ and thresholds $\frac{1}{|U|}$.
	\end{itemize}
	
	Notice the number of vertices and edges in $G'$:
	$$|U_1 \cup U_2| = |U| + |E| + 2\left( \frac{|U|^2 - |U|}{2} - |E| \right) = |U|^2 - |E|,$$
	$$|E'| = |U|^2 - |U|.$$
	
	Set the desired penetration rate in $G'$ to $\zeta = \frac{k |U|}{|U|^2 - |E|}$ (this is the fraction of nodes we want to reverse the defaults of in the economic network). Notice that
	$$\zeta |U_1 \cup U_2| = \frac{k |U|}{|U|^2 - |E|} |U|^2 - |E| = k|U|,$$
	which will be the desired penetration in the reduction graph to correspond to the independent set (which we prove below).

	\paragraph{Gadget is instance of economic network intervention.}
	We now show that the independent set problem on $G'$ translates to an instance $(C,\beta,\bm\theta,D,\mathbf p)$ of the economic network intervention problem. Let $A$ be the adjacency matrix of $G'$. Since $G'$ is a 2-layer DAG, we have $A^t = 0$ for integers $t>1$. Then the Neumann series is
	$$(I-A)^{-1} = I+A.$$
	
	Notice that $A$ is non-negative column-substochastic with zero diagonal. Thus we take $C = A$, and $\hat C$ is well-defined.
	
	\vspace{0.25cm}
	\noindent\textbf{Claim:}
	$(\beta,\bm\theta,D,\mathbf p)$ can be chosen such that, before intervention, all nodes fail with end values $\mathbf v=0$, $\tilde\theta_u = \frac{1}{|U|}$ for all $u$, and $\beta \geq 1$.
	
	\vspace{0.25cm}
	\noindent\textbf{Proof of claim:}
	To find such a $(\beta,\bm\theta,D,\mathbf p)$, we can setup the following system
	$$\begin{aligned}
	\mathbf V &= (I+C)(D\mathbf p - \beta \Ind_{U_1\cup U_2}) = 0 \\
	\theta_u &> \Big[\hat C (I+C)D\mathbf p \Big]_u \text{ for } u\in U_1 \\
	\theta_u &> \Big[\hat C (I+C)(D\mathbf p - \beta \Ind_{U_1}) \Big]_u \text{ for } u\in U_2 \\
	\tilde\theta_u &= \Big[\hat C^{-1}\bm\theta - (I+C)D\mathbf p - C\beta \Ind_{U_1\cup U_2 \setminus \{u\}} \Big]_u = \frac{1}{|U|} \text{ for all } u \\
	\beta &\geq 1.
	\end{aligned}$$
	The system has the same number of variables as dimensions. Because of the 2-layer DAG structure, it is simple to see that the system is solvable.
	
	Notice that in the equation for $\tilde\theta$ is valid. Taking failure set $T$, we have
	$$\begin{aligned}
	\tilde\theta_u &= \Big[ \hat C^{-1}\bm\theta - (I+C)(D\mathbf p - \beta \Ind_{T\setminus\{u\}}) \Big]_u \\
	&= \Big[ \hat C^{-1}\bm\theta - (I+C)D\mathbf p - C\beta \Ind_{T\setminus \{u\}} \Big]_u
	\end{aligned}$$
	because $[I\beta \Ind_{T\setminus\{u\}}]_u = 0$.

	\vspace{0.25cm}
	\noindent\textbf{Claim:}
	The effect of reversing defaults $S$ propagates to other nodes through $f(S) = C\beta\Ind_S$.
	
	\vspace{0.25cm}
	\noindent\textbf{Proof of claim:}
	First notice that for all nodes $u$,
	$$\Big[ (I+C)\beta \Ind_u \Big]_u = \beta_u.$$
	This is a simple result because $C$ has zero diagonal and the only nonzero entry of $\Ind_u$ is the $u$th entry; thus there is $0$ contribution from $C\beta\Ind_u$ for the $u$th entry.
	
	Then we have
	$$\begin{aligned}
	f(S) &= (I+C)\beta \Ind_S - \sum_{u\in S} I_u(I+C)\beta \Ind_u \\
	&= (I+C)\beta \Ind_S - \sum_{u\in S} I_u \beta \Ind_u \\
	&= (I+C)\beta \Ind_S - \beta \Ind_S \\
	&= C\beta \Ind_S.
	\end{aligned}$$

	\vspace{0.25cm}
	\noindent\textbf{Claim:} If we reverse the default of a node in $U_1$, then its neighbors in $U_2$ are also saved from default.
	
	\vspace{0.25cm}
	\noindent\textbf{Proof of claim:}
	Suppose we reverse the default of $u\in U_1$. Suppose $w\in U_2$ is a neighbor of $u$. Then $w$'s value is affected by
	$$[f(u)]_w = [C\beta \Ind_u]_w
	= \frac{\beta}{|U|} \\
	> \frac{1}{|U|} \\
	= \tilde\theta_w$$
	since $\beta \geq 1$. Thus $w$'s default is also reversed.

	\vspace{0.25cm}
	To complete the translation into the economic network intervention problem, define the following:
	$$b = \frac{k}{|U|}$$
	$$\alpha = 1$$
	$$\ell = 1-\zeta.$$

	In intuitive terms, the corresponding economic network is a 2-layer DAG, in which the only cross-holdings are the shares in the first layer held by the second layer. In this case, the interactions are quite simple, described solely by $C$. In this network, every node starts in default. We can pay $\tilde\theta=\frac{1}{|U|}$ to reverse a node's default. Our budget is $b$ and we can choose at most $k$ nodes to intervene on.
	
	
	

	\paragraph{Reduction to integral case.}
	We first consider the integral case and then extend to the fractional case. We want to select a subset $S$ of $k$ nodes from $G'$ such that, if we provide payments equal to their $\tilde\theta$, a cascade of reverse-defaults occurs of size at least $\zeta |U_1 \cup U_2|$ (i.e., at most $\ell |U_1 \cup U_2|$ nodes fail after intervention). This occurs if and only if $G$ has an independent set of size $k$, as we prove next.
	
	First, note that sets $S \subseteq U_1$ always dominate sets $S \subseteq U_1 \cup U_2$ with $S \subsetneq U_1$. This is because, by construction, reversing the default of any node in $U_1$ in turn impacts its neighbors in $U_2$, reversing their defaults, whereas reversing the default of a node in $U_2$ does not impact its neighbors in $U_1$. Since each node in $U_2$ has a neighbor in $U_1$, it always makes sense to impact such a neighbor instead of the considered node in $U_2$. Thus it is sufficient to consider only solutions in $U_1$. Notice that this extends to the fractional case since threshold-crossing payments are of the same size for nodes in $U_1$ and $U_2$.
	
	Each node in $U_1$ has $|U|-1$ neighbors, and two nodes in $V_1$ share a neighbor if and only if they are neighbors in $G$. So if we pick the subset $S\subseteq U_1$, the size of the default reverse cascade is
	$$\# \text{default reverses} = |U| |S| - |\{ \{i,j\} \in E | i,j \in S \}|.$$
	E.g., if no nodes in $S$ are connected in $G$, then the second term is $0$ and each default reverting node impacts itself and $|U|-1$ unique nodes in $U_2$ for a total of $|S| + (|U|-1)|S| = |U| |S|$ nodes.
	
	The number of default reversals is $\geq \ell |U_1 \cup U_2| = k|U|$ if and only if $\forall u,v \in S$, $\{u,v\} \notin E$, which is that case if and only if there is an independent set of size $k$ in $G$.
	
	\paragraph{Reduction to fractional case.}
	Notice that this easily extends to the fractional case. In this case, we want to find payments such that $\sum_i \gamma_i \leq b = \frac{k}{|U|}$ and we save $\zeta |U_1 \cup U_2|$ nodes from failure (i.e., at most $\ell |U_1 \cup U_2|$ nodes fail after intervention). In $G'$, all edges and thresholds have value $\frac{1}{|U|}$. Given the structure of $G'$, optimal node payments will obey $\gamma_i \in \{0, \frac{1}{|U|} \}$. This is because a payment to a node in $U_1$ is again always better than a payment to a node in $U_2$ (same argument as before), and any payment smaller than $\frac{1}{|U|}$ will result in no default reversals in $U_1$, and hence no subsequent effect on $U_2$. Thus there is one-to-one correspondence between optimal integral solutions and optimal fractional solutions. Thus the fractional case is NP-hard in general.
\end{proof}

\noindent\rule{\textwidth}{1pt}
\paragraph{Proposition~\ref{result:f_submodular}} \hypertarget{pf:f_submodular}{}
\begin{proof}
	To simplify notation, define $A:=(I-C)^{-1}\beta$.
	
	(Monotonicity) Let $T\subset U$ and $u\in U\setminus T$. Then we have
	$$\begin{aligned}
	f(T\cup \{u\})
	&= A\Ind_{T\cup \{u\}} - \sum_{j\in T\cup \{u\}} I_j A \Ind_j \\
	&= A\Ind_T - \sum_{j\in T} I_j A \Ind_j + A \Ind_u - I_u A \Ind_u \\
	&= f(T) + A \Ind_u - I_u A \Ind_u.
	\end{aligned}$$
	Since $A$ is non-negative, the second term is $\geq 0$. The third term only affects the $u$th component, and then only cancels the contribution of the second term. Thus we have $f(T\cup \{u\}) \geq f(T)$.
	
	(Submodularity) Let $S \subseteq T \subseteq U$ and $u\in U \setminus T$. From the above equations, we have
	$f(T\cup \{u\}) - f(T) = A \Ind_u - I_u A \Ind_u$.
	and similarly with $S$. Thus the submodularity condition
	$f(S\cup \{u\}) - f(S) \geq f(T \cup \{u\}) - f(T)$
	holds with equivalence.
\end{proof}

\noindent\rule{\textwidth}{1pt}
\paragraph{Theorem~\ref{result:intervention_submodular}} \hypertarget{pf:intervention_submodular}{}
\begin{proof}
	Recall that the intervention problem can be expressed in an influence maximization-like form. By assumption, $w$ is normalized, monotone, and submodular, and $f$ is normalized. And by Prop.~\ref{result:f_submodular}, $f$ is monotone and submodular. Notice that the intervention problem is easily normalized (in a different sense) so as to restrict each $f_i$ and $\tilde \theta_i$ to the range $[0,1]$. Then by Theorem~1 in \cite{mossel07}, the integral intervention problem has $\sigma(S_0)$ normalized, monotone, and submodular. And by Theorems~2-3 in \cite{demaine14}, the fractional intervention problem has $\sigma(\bm \gamma)$ normalized, monotone, and submodular (note that these definitions are modified to describe non-set functions in the fractional case).
\end{proof}

\noindent\rule{\textwidth}{1pt}
\paragraph{Corollary~\ref{result:greedy_approx}} \hypertarget{pf:greedy_approx}{}
\begin{proof}
	This follows using the same application of results as in \cite{kempe03}. In particular, the results of \cite{nemhauser77},\cite{nemhauser78} show that a greedy hill-climbing algorithm approximates the optimum of monotone submodular problems to within a factor of $(1-1/e)$. Given that $\sigma$ has to be approximated, the result can be extended to show that for any $\epsilon>0$, there is $\delta>0$ such that by using $(1+\delta)$-approximate values for the $\sigma$ function, we obtain a $(1-1/e-\epsilon)$-approximation. For the fractional case, this uses Theorem~4 in \cite{demaine14}.
\end{proof}

\noindent\rule{\textwidth}{1pt}
\paragraph{Theorem~\ref{result:max_shock_hard}} \hypertarget{pf:max_shock_hard}{}
\begin{proof}
	First consider a specific subclass of economic network instances. We will reduce independent set to an instance of this subclass. The subclass has the following properties:
	\begin{itemize}
		\item Asset prices $\mathbf  p$ take values in $\{0,1\}$.
		\item $D$ is row-sub-stochastic, such that a firm's underlying assets can be valued at most 1.
		\item $C = 0$, in which case $\hat C = I$ and $(I-C)^{-1}=I$.
		\item $\beta=0$, in which case a firm's value is in $[0,1]$.
		\item $b$ is an integer.
	\end{itemize}
	As a result, the shock to be chosen in our problem, if applied to asset $i$, can change it's price from $1$ to $0$. The problem at hand is now to find a set of $b$ assets that, if set to 0, cause $\ell |U|$ firms to default.
	
	Next consider a reformulation of the network process into a bipartite graph $G'$ as follows:
	\begin{itemize}
		\item Add nodes for each underlying asset. Denote these nodes $U_1$.
		\item Add nodes for each firm. Denote these nodes $U_2$.
		\item For each $u \in U_1$, add a weighted directed edge from $u$ to nodes in $U_2$ according to the matrix $D$. The weights here represent the effect of the asset on the book values of firms that own those assets in the simple setting with $C=0$.
	\end{itemize}
	Assume the assets in $U_1$ are initially set to 1. If an asset is changed to 0, (negative) impact is exerted on its connections in $U_2$ via $D$, lowering those firms' values. If enough (negative) impact is exerted on a firm in $U_2$, its value decreases below threshold, triggering default. The equivalent problem is to find a set of $b$ nodes in $U_1$ such that, if set to 0, cause $\ell |U|$ firms to default. 
	
	To reduce from independent set, we can follow essentially the same reduction as in Theorem~\ref{result:interventions_hard} to a process on a bipartite graph like above. With appropriate definition of parameters, this is an instance of the subclass of economic networks above. And thus independent set reduces to economic network maximum shock problem.
\end{proof}



\section{Algorithms}\label{sec:algorithms}
We provide explicit descriptions of the optimal intervention approximation algorithms to aid the reader, as their adaptations in the influence maximization literature are usually not made explicit. The algorithms below use the following problem setting consistent with the intervention problem developed in the paper:
\begin{itemize}
	\item $f(S)$ outputs the intervention impact vector exerted by set $S$ on each node.
	\item $w(S)$ outputs a weight of node set $S$.
	\item $\Theta$ is node threshold distribution, uniformly distributed between $\bm\tilde\theta_{\min}$ and $\bm\tilde\theta_{\max}$. The thresholds $\bm\tilde\theta$ are sampled from this distribution.
	\item $b$ = budget.
\end{itemize}

There are three primary intervention algorithms. The remaining algorithms serve as helper functions used in these primary algorithms.
\begin{itemize}
    \item Algorithm~\ref{alg:int_greedy} is the greedy algorithm for approximating optimal integral interventions with 63\% guarantees.
    \item Algorithm~\ref{alg:frac_greedy} is the greedy algorithm for approximating optimal fractional interventions with 63\% guarantees.
\end{itemize}
Notice that these algorithms need to re-estimate a $\hat\sigma$ high-dimensional integral at each step through Monte Carlo, which is often too computationally intense to run in high-dimensional systems, even though it is technically polynomial time with the Monte Carlo capped at a constant factor.
\begin{itemize}
    \item Algorithm~\ref{alg:discount_frac} is a fast heuristic greedy algorithm that is very close to the ideal fractional greedy algorithm. It does not come with provable guarantees, but is used similarly in influence maximization with large success.
\end{itemize}
Full and optimized Python implementation is available at \url{https://github.com/aklamun/optimal_intervention}.

\begin{algorithm}[H]
	\algorithmicrequire { set $S$, set function $f$, thresholds $\bm\tilde\theta$}
	\begin{algorithmic}
		\State Initialize $S_0 \leftarrow \emptyset$, $S_1 \leftarrow S$, $i \leftarrow 1$
		\While {$S_i \neq S_{i-1}$}
		\State $S_{i+1} = \{\text{node } v | f(S_i)[v] \geq \bm\tilde\theta[v]\}\cup S_i$
		\State $i \leftarrow i+1$
		\EndWhile
		\State \Return $S_i$
	\end{algorithmic}
	\caption{$\texttt{CalcIntCascade}(S;f,\bm\tilde\theta)$} \label{alg:calc_int_cascade}
\end{algorithm}

\begin{algorithm}[H]
	\algorithmicrequire { set $S$, set function $f$, weight function $w$, thresholds distr. $\Theta$, sample size $k=1e4$}
	\begin{algorithmic}
		\State Initialize $\sigma \leftarrow 0$
		\For {$i \leq k$}
		\State Sample $\bm\tilde\theta \sim \Theta$
		\State $T, = \texttt{CalcIntCascade}\Big( S; f, \bm\tilde\theta \Big)$
		\State $\sigma \leftarrow \sigma + w(T)$
		\EndFor
		\State \Return $\sigma/k$
	\end{algorithmic}
	\caption{$\hat\sigma(S)$ estimate of $\sigma(S)$ for integral intervention} \label{alg:sigma}
\end{algorithm}

\begin{algorithm}[H]
	\algorithmicrequire { set function $f$, weight function $w$, thresholds distr. $\Theta$, budget $b$}
	\begin{algorithmic}
		\State Initialize $S_0 \leftarrow \emptyset$, $i\leftarrow 0$
		\While {$|S_i| < b$}
		\For {node $v \notin S_i$}
		\State $\mathbf{q}[v] = \hat\sigma\Big(S_i \cup \{v\}; f, \Theta, w \Big)$
		\EndFor
		\State $S_{i+1} \leftarrow S_{i}\cup \{\arg\max \mathbf{q}\}$, $i\leftarrow i + 1$
		\EndWhile
		\If {$|S_i| \leq b$}
		\State \Return $S_i$
		\Else
		\State \Return $S_{i-1}$
		\EndIf
	\end{algorithmic}
	\caption{\texttt{GreedyInt} = Greedy algorithm for optimal integral intervention} \label{alg:int_greedy}
\end{algorithm}


\begin{algorithm}[H]
	\algorithmicrequire { vector $\bm \gamma$, set function $f$, thresholds $\bm\tilde\theta$}
	\begin{algorithmic}
		\State Initialize $S_0 \leftarrow \emptyset$, $i\leftarrow 1$
		\State $S_1 \leftarrow \{\text{node } v | \bm \gamma_v \geq \bm\tilde\theta_v\}$
		\While {$S_i \neq S_{i-1}$}
		\State $S_{i+1} = \{\text{node } v | f(S_i)[v] + \bm \gamma_v \geq \bm\tilde\theta_v\}$
		\State $i \leftarrow i+1$
		\EndWhile
		\State \Return $S_i$
	\end{algorithmic}
	\caption{$\texttt{CalcFracCascade}(\bm \gamma;f,\bm\tilde\theta)$} \label{alg:calc_frac_cascade}
\end{algorithm}

\begin{algorithm}[H]
	\algorithmicrequire { vector $\bm\gamma$, set function $f$, weight function $w$, thresholds distr. $\Theta$, sample size $k=1e4$}
	\begin{algorithmic}
		\State Initialize $\sigma \leftarrow 0$
		\For {$i \leq k$}
		\State Sample $\bm\tilde\theta \sim \Theta$
		\State $T = \texttt{CalcFracCascade}\Big( \bm\gamma; f, \bm\tilde\theta \Big)$
		\State $\sigma \leftarrow \sigma + w(T)$
		\EndFor
		\State \Return $\sigma/k$
	\end{algorithmic}
	\caption{$\hat\sigma(\bm\gamma)$ estimate of $\sigma(\bm\gamma)$ for fractional intervention} \label{alg:sigma_frac}
\end{algorithm}

\begin{algorithm}[H]
	\algorithmicrequire { set function $f$, weight function $w$, thresholds distr. $\Theta$, budget $b$}
	\begin{algorithmic}
		\State Initialize $\mathbf{\bm\gamma_0} \leftarrow \mathbf{0}$, $i\leftarrow 0$
		\While {$\mathbf{1}^T\mathbf{\bm\gamma_i} < b$}
		\State $S_i = \{\text{node } v | \mathbf{\bm\gamma_i}[v] > 0\}$
		\For {node $v \notin S_i$}
		\State $\mathbf{\bm\gamma_v} = \mathbf{\bm\gamma_i} + \Big( \theta_{\max}[v] - \Gamma^+(v,S_i)\Big) \mathbf{1}_{v}$
		\State $\mathbf{q}[v] = \hat\sigma\Big(\mathbf{\bm\gamma_v}; f, \Theta, w \Big)$
		\EndFor
		\State $u = \arg\max \mathbf{q}$
		\State $\mathbf{\bm\gamma_{i+1}} \leftarrow \mathbf{\bm\gamma_i} + \Big( \bm\tilde\theta_{\max}[u] - \Gamma^+(u,S_i)\Big) \mathbf{1}_{u}$, $i\leftarrow i + 1$
		\EndWhile
		\If {$\mathbf{1}^T\mathbf{\bm\gamma_i} \leq b$}
		\State \Return $\mathbf{\bm\gamma_i}$
		\Else
		\State \Return $\mathbf{\bm\gamma_{i-1}}$
		\EndIf
	\end{algorithmic}
	\caption{\texttt{GreedyFrac} = Greedy algorithm for optimal fractional intervention} \label{alg:frac_greedy}
\end{algorithm}

\begin{algorithm}[H]
	\algorithmicrequire { set $A$, set function $f$, node $v$}
	\begin{algorithmic}
		\State \Return $f(A)[v]$
	\end{algorithmic}
	\caption{$\Gamma^+(v,A)$ = total sum of weight of edges from set $A$ to node $v$} \label{alg:gamma_plus}
\end{algorithm}

\begin{algorithm}[H]
	\algorithmicrequire { set $A$, set function $f$, node $v$}
	\begin{algorithmic}
		\State \Return $\mathbf{1}_A^T f(\{v\})$
	\end{algorithmic}
	\caption{$\Gamma^-(v,A)$ = total sum of weight of edges from node $v$ to set $A$} \label{alg:gamma_neg}
\end{algorithm}

\begin{algorithm}[H]
	\algorithmicrequire { set function $f$, weight function $w$, thresholds distr. $\Theta$, budget $b$}
	\begin{algorithmic}
		\State Initialize $\mathbf{x_0} \leftarrow \mathbf{0}$, $i\leftarrow 0$
		\While {$\mathbf{1}^T\mathbf{\bm\gamma_i} < b$}
		\State $S_i = \{\text{node } v | \mathbf{\bm\gamma_i}[v] > 0\}$
		\For {node $v \notin S_i$}
		\State $\mathbf{q}[v] = \Gamma^-(v, V\backslash S_i)$
		\EndFor
		\State $u = \arg\max \mathbf{q}$
		\State $\mathbf{\bm\gamma_{i+1}} \leftarrow \mathbf{\bm\gamma_i} + \Big( \bm\tilde\theta_{\max}[u] - \Gamma^+(u,S_i)\Big) \mathbf{1}_{u}$, $i\leftarrow i + 1$
		\EndWhile
		\If {$\mathbf{1}^T\mathbf{\bm\gamma_i} \leq b$}
		\State \Return $\mathbf{\bm\gamma_i}$
		\Else
		\State \Return $\mathbf{\bm\gamma_{i-1}}$
		\EndIf
	\end{algorithmic}
	\caption{\texttt{DiscountFrac} heuristic intervention algorithm} \label{alg:discount_frac}
\end{algorithm}

\end{document}